\documentstyle[12pt]{article}
\newcommand{\eq}{\begin{equation}}
\newcommand{\ee}{\end{equation}}
\newcommand{\eqa}{\begin{eqnarray}}
\newcommand{\eea}{\end{eqnarray}}

\newcommand{\s}{{\sigma}}
\newcommand{\psib}{{\bar{\psi}}}
\newcommand{\bb}{{\bar{B}}}
\newcommand{\w}{{\omega}}

\newcommand{\ap}{{\alpha}}

\newcommand{\bt}{{\beta}}

\begin{document}
\vspace{0.2in}
\begin{center}
\large\bf
The Effects of Electron-Electron Interactions on the Integer Quantum
Hall Transitions \\
\vspace{0.5in}
\normalsize\rm
Dung-Hai Lee$^{(a)}$ and Ziqiang Wang$^{(b)}$ \\
\vspace{0.2in}
\em
$^{(a)}$Department of Physics, University of California at Berkeley, 
Berkeley, CA 94720 \\
\vspace{0.2in}
$^{(b)}$Department of Physics, Boston University, Boston, MA 02215 \\

\vspace{1.0in}
\bf
Abstract
\end{center}
\vspace{0.2in}
\vbox{\narrower\smallskip
\rm
We study the effects of electron-electron interaction on the critical 
properties of the plateau transitions in the {\it integer} quantum Hall 
effect. We find the renormalization group dimension associated with 
short-range interactions to be $-0.66\pm0.04$. 
Thus the non-interacting fixed point (characterized $z=2$ and
$\nu\approx 2.3$) is stable. For the Coulomb interaction, 
we find the correlation effect is a marginal perturbation at a
Hartree-Fock fixed point ($z=1$, $\nu\approx 2.3$) by dimension
counting. Further calculations are needed to determine its stability 
upon loop corrections.
\smallskip}
\vskip 1cm
\begin{verbatim}
PACS numbers: 73.50.Jt, 05.30.-d, 74.20.-z
\end{verbatim}
\newpage
\setlength{\baselineskip}{.375in}
\rm
The plateau transitions has been one of the unsolved
problems in the quantum Hall effect. The fundamental questions that remain
unanswered are a) what are the effects of electron-electron
interaction on the integer transitions. b) What are the effects of the 
quasiparticle statistics on the fractional transitions.
Almost all recent works on the integer plateau transitions 
are based on numerical analyses of models of {\it non-interacting} 
electrons \cite{review}. An important outcome of the these efforts is the
consensus on the approximate values of the localization length 
exponent ($\nu$), and several others characterizing 
the participation ratio and its higher moments \cite{review}.
In particular, the result $\nu\approx 2.3$ is in excellent agreement with the 
measured value \cite{wei,koch}.
Nonetheless, such basic issues as the relevance and the effects of
the electron-electron interaction have not been addressed.  
The necessity to understand the interaction effects becomes even more 
pressing after recent experimental reports of the dynamical exponent
$z=1$, instead of the non-interacting 
value $2$ \cite{wei2,engel}.
In this paper, we focus on the effects of the interactions on 
the {\it integer} plateau transitions. 

Our strategy is the following. 
We take the {\it 2+1} dimensional non-interacting theory as the 
starting point. We then ask what the effects are of turning on 
the interaction. In practice, we calculate the renormalization group  
(RG) dimension of the interaction Hamiltonian, and also look at the
other possible interactions it generates upon renormalization. 
This is a standard exercise when one analyzes the stability of a known 
critical point. However, unlike many cases where one has {\it analytic} 
knowledge of the critical point in question, in the present case 
such knowledge is lacking. Thus the results that we report in this letter 
are based on {\it numerical} calculations of various correlation
functions at the non-interacting/Hartree-Fock fixed point.

For {\it short-range} interaction we obtained its RG dimension
$\approx -0.65$, and found that to the second order in the interaction 
strength no relevant operators are generated upon renormalization. 
Thus we conclude that short-range interaction is an irrelevant perturbation 
at the non-interacting fixed point. Hence for screened electron-electron 
interactions $z=2$ and $\nu\approx 2.3$.
For Coulomb interaction,
we find that it is relevant at the 
{\it non-interacting} fixed point. 
However, by dimension counting, 
we find that due to a linear suppression in the density of states (DOS)
\cite{mac}, the correlation effect is only a {\it marginal} 
perturbation at the Hartree-Fock fixed point. For the latter, 
we find $z=1$ and $\nu\approx 2.3$. The Hartree-Fock fixed point provides  
a concrete example where Coulomb interaction modifies the 
dynamical exponent and not the static one.
The root of such behavior is the {\it non-critical} 
suppression of the DOS. Indeed, as was shown in 
Ref.\cite{mac}, the Hartree-Fock DOS vanishes 
linearly with $|E-E_F|$ {\it regardless of whether the Fermi energy
$E_F$ coincides with the critical value}. This suppression resulted 
in a change of $z$ from 2 to 1, and a degradation of the RG dimension 
of the {\it residual} Coulomb Hamiltonian from 1 to 0. 
We do not yet know the effects of 
the residual Coulomb interaction upon further loop corrections.

We start with non-interacting electrons described by the following 
Euclidean action: (in units $e/c=\hbar=k_B=1$.)
\eq
S_0=\int d^2x\sum_{\w_n}\psib_{\w_n}(x)[-i\w_n+\Pi^2+V_{\rm imp}(x)]
\psi_{\w_n}(x).
\ee
In the above, $\psi$ is the fermion Grassmann field, $\w_n=(2n+1)\pi/\bt$
is the fermion Matsubara frequency, $V_{\rm imp}$ is the disorder
potential, and $\Pi^2\equiv -{1\over {2m}}\sum_k(\partial_k-iA_k(x))^2$ 
where $A_k(x)$ is the external vector potential. 
The action describing the interaction reads
\eq
S_{\rm int}=T\sum_{\w_1,...,\w_4}\delta_{\w_1+\w_2,\w_3+\w_4}\int d^2xd^2y
V(\vert x-y\vert)\psib_{\w_1}(x)\psib_{\w_2}(y)\psi_{\w_4}(y)\psi_{w_3}(x).
\ee
In this paper we consider $V(|x-y|)=g/|x-y|^\lambda$.
The total action is $S=S_0+S_{\rm int}$, in which
$S_{\rm int}$ couples the otherwise 
independent frequency components of $S_0$ together.
To emphasize the symmetry property of $S_{\rm int}$ we rewrite it as
\eq
S_{\rm int}={T\over 4}\sum_{\w_1,...,\w_4} \delta_{\w_1+\w_2,\w_3+\w_4} 
\int d^2x d^2y V(\vert x-y\vert) 
\bb_{(\w_1,\w_2)}(x,y)B_{(\w_3,\w_4)}(y,x),
\ee
where 
$
\bb_{(\w_1,\w_2)}(x,y)\equiv \psib_{\w_1}(x)\psib_{\w_2}(y)+
\psib_{\w_2}(x)\psib_{\w_1}(y).
$
In the following we imagine sitting at the fixed point of $S_0$ and 
numerically analyze the scaling properties of  $<S_{\rm int}>$ and 
$<S_{\rm int}^2>$ in finite periodic systems of linear dimension $L$ 
and imaginary time dimension $1/T$. (Hereafter $<...>$ denotes quantum 
and impurity averages.)
Since we are after the {\it universal} scaling properties of 
various correlation functions, any representation of $S_0$ 
which produces the right universality class suffices. 
In the following we choose the ``quantum percolation'' 
(or the network) model of Chalker and Coddington. 
For details about this model the readers are referred 
to Ref.\cite{cc}. Moreover, the numerical calculations reported 
here are done using the U(2n)$\vert_{n\to0}$ Hubbard model
representation of the network model \cite{lw}.

In order to evaluate $<S_{\rm int}>$, it is necessary to know the 
correlation function 
$<\bb_{(\w_1,\w_2)}(x,y)B_{(\w_3,\w_4)}(y,x)> = 
\delta_{\w_1,\w_3}\delta_{\w_2,\w_4} \Gamma'^{(4)}(x,y;\w_1,\w_2,L)$. 
To extract the critical piece of $\Gamma'^{(4)}$,
it is important to perform the trace-decomposition \cite{wegner,bk}.
Thus we write
\eq
\Gamma'^{(4)}(x,y;\w_1,\w_2,L)\equiv\Gamma^{(4)}(x,y;\w_1,\w_2,L)+
<\psib_{\w_1}(x)\psi_{\w_1}(x)\psib_{\w_1}(y)\psi_{\w_1}(y) +
(\omega_1\to\omega_2)>.
\ee
Each of the last two terms in Eq.~(4) involves only a single Matsubara 
frequency and is non-critical.
$\Gamma^{(4)}$ is only critical when  $\w_1\w_2<0$ \cite{lw1}.
Since the scaling dimension of $\psib_{\w_1}(x)\psi_{\w_2}(x)$ 
is zero at the non-interacting fixed point 
(i.e. the density of states has no anomalous dimension) 
\cite{lw}, $\Gamma^{(4)}$ obeys the following scaling form
\eq
\Gamma^{(4)}(x,y;\w_1,\w_2,L)=
{\cal F}_1({{\vert x-y\vert}\over L},\w_1 L^2,\w_2 L^2).
\ee 
Note that near the noninteracting fixed point, it is sufficient to
consider any pair of a positive and a negative frequency.
The latter scales with $L^{-2}$ in Eq.~(5) which reflects the fact that
$z=2$ at the noninteracting fixed point.
We have calculated $\Gamma^{(4)}$
using the Monte Carlo method of Ref.~\cite{lw} for $\w_{1,2}L^2=
{\rm const}_{1,2}$. The details of the calculation will be reported
elsewhere \cite{lw1}.
The scaling behavior of $\Gamma^{(4)}$ versus $\vert x-y\vert/L$ is
shown in Figure 1. 
The results are consistent with
${\cal F}_1({{\vert x-y\vert}\over L},\w_1 L^2,\w_2 L^2)\sim 
\left({{\vert x-y\vert}\over L}\right)^{x_{4s}}$
for $\vert x-y\vert/L<<1,\vert(\w_{1,2}L^2)\vert^{-1/2}$,
and $x_{4s}\approx 0.65$.
Thus, in terms of the properly scaled variables,
the first order correction to the {\it singular part} of the
quench-averaged action is,
\eq
\Delta S^{(1)}_{\rm sing}={{TL^2}\over 4}\left({g\over {L^{\lambda-2}}}\right)
 {\sum_{n_1,n_2}}'\int\! d^2\left({x\over
   L}\right)\int\!d^2\left({y\over L}\right)
  \left\vert{L\over x-y}\right\vert^{\lambda}
 {\cal F}_1\left({{\vert x-y\vert}\over L};\pi(2n_{1,2}+1)TL^2\right).
\ee
In the above $\sum'$ denotes the restricted sum satisfying $n_1n_2<0$.
Let us change the integration variables $d^2(x/L)d^2(y/L)$ to 
$d^2(x+y)d^2(x-y)/L^2$. The part that depends on the relative coordinate reads
\eq
\int d^2\left({{x-y}\over L}\right)\left({L\over {\vert x-y\vert}}
\right)^{\lambda}
{\cal F}_1\left({{\vert x-y\vert}\over L},\w_1 L^2,\w_2 L^2\right),
\ee 
where the upper limit of the integral is $1$ and the lower one is $a/L$ 
($a$ is the lattice spacing). Naively, one would deduce from
Eq.~(6) that the RG dimension of $g$ is $2-\lambda$ as the result of 
dimensional analyses. 
This conclusion can be modified if the integral in Eq.~(7) depends on
$a/L$, i.e. if it diverges at the lower limit. Since 
${\cal F}_1\sim (\vert x-y\vert/L)^{x_{4s}}$
for $\vert x-y\vert/L \ll 1$,
the integral diverges (we will henceforth refer to this case as that of 
short-range interaction) 
when $\lambda\ge x_{4s}+2$, and converges (long-range interaction) otherwise.

Let us now concentrate on the case $\lambda >x_{4s}+2$ 
(i.e. short-range interaction). Simple analyses
of Eq.~(6) show that
\eq
\Delta S^{(1)}_{\rm sing}=\left({g\over {L^{\lambda-2}}}\right)\left[A+
B\left({a\over L}\right)^{2+x_{4s}-\lambda}\right],
\ee
where $A$ and $B$ are non-universal functions of $TL^2$.
Since $\lambda-2>x_{4s}$, the asymptotic scaling behavior of 
$\Delta S^{(1)}_{\rm sing}$ is controlled by
$\Delta S^{(1)}_{\rm sing}=B{{u}/ {L^{x_{4s}}}}$
where $u\equiv ga^{2+x_{4s}-\lambda}$. In the language of the renormalization
group, the density operators at nearby points have fused
together to form a new operator with a RG dimension $-x_{4s}$. 
Thus for screened Coulomb interactions, 
we conclude that the RG dimension for $T$ is $2$ (thus $z=2$), and that
for $u$ is $-x_{4s}\approx -0.65$.
Therefore to this order the interaction is {\it irrelevant}.
Here we note that if similar analyses are 
done for the weak-field (i.e. the 
``singlet only'') metal-insulator transition in 
$2+\epsilon$ dimensions, one obtains
$x_{4s}=\sqrt{2\epsilon}$ agreeing with the results obtained in, {\it e.g.}, 
Ref.\cite{bk}.  

In order to perform a self-consistency check on the RG dimension of $u$,
and to study the fusion products \cite{cardy} of two interaction
operators, we next calculate $<S_{\rm int}^2>$.
For that purpose we need to consider
$<\bb_{(\w_1,\w_2)}(x,y)B_{(\w_3,\w_4)}(y,x)\bb_{(\w_5,\w_6)}
(x',y')B_{(\w_7,\w_8)}(y',x')>_c=\delta_{\w_1,\w_7}\delta_{\w_2,\w_8}
\delta_{\w_3,\w_5}\delta_{\w_4,\w_6}\Gamma^{(8)}(x,y,x',y'; 
\w_1,\w_2,\w_3,\w_4;L)$.
For short-range interactions we only need to concentrate on the limit 
$\vert x-y\vert,\vert x'-y'\vert<<\vert R-R'\vert$, where $R=(x+y)/2$ 
and $R'=(x'+y')/2$. In that limit and for $\w_1\w_2<0, \w_3\w_4<0$, 
(other combinations give non-critical contributions \cite{lw1}), the result is
\eq
\Gamma^{(8)}(x,y,x',y';\{\w_i\};L)=
{\cal F}_2\left(\left\vert {x-x'\over R-R'}\right\vert,\left\vert 
{y-y'\over R-R'}\right\vert,{\vert R-R'\vert\over L},\{\w_iL^2\}\right).
\ee
In the limit $\vert x(y)-x'(y')\vert/\vert
R-R'\vert<<1,\vert\w_iL^2\vert^{-1/2}$,
${\cal F}_2$ reduces to
\eq
{\cal F}_2\sim\left\vert{x-y\over R-R'}\right\vert^{x_{4s}}
\left\vert {x'-y'\over R-R'}\right\vert^{x_{4s}}{\cal F}_3
(\vert R-R'\vert/L,\{\w_iL^2\}).
\ee
The result for $\vert R-R'\vert^{2x_{4s}}\Gamma^{(8)}$ versus 
$\vert R-R'\vert/L$ for small, typical, 
fixed $\vert x-y\vert$, $\vert x'-y'\vert$, $\w_i={\cal O}(1/L^2)$, 
and $x_{4s}=0.65$, is shown in Figure 2. 
This result indicates that the previously obtained $x_{4s}\approx 0.65$
is the consistent scaling dimension of the the short-range interaction.
Going through similar manipulations one can show that the second order 
correlation correction to the singular part of the 
quench-averaged action, $\Delta S^{(2)}_{\rm sing}$, is
\eq
\Delta S^{(2)}_{\rm sing}=-(TL^2)^2 {{u^2}\over {L^{2x_{4s}}}}
{\sum_{n_1,...,n_4}}'
\int\! d^2\left({R\over L}\right)d^2\left({{R'}\over L}\right)\left\vert 
{L\over {R-R'}}\right\vert^{2x_{4s}}{\cal F}_4\left[{\vert R-R'\vert\over
  L},\pi(2n_{1\to4}+1)TL^2\right],
\ee
where $\sum'$ denotes the restricted sum satisfying $n_1n_2<0$ and
$n_3n_4<0$, and $F_4\propto F_3$.
In Eq.~(11) let us convert $d^2Rd^2R'$ to $d^2(R+R')d^2(R-R')$. 
In the integral over the relative coordinates, the short-distance 
cutoff is again $a/L$. As before, new dependence on $L$ could emerge if the 
integral over $R-R'$ diverges at the lower limit. In general, if
${\cal F}_4(\vert R-R'\vert /L,\{\w_iL^2\})\sim\vert{{R-R'}\over L}
\vert^{\ap}$, and if $2x_{4s}-\ap-2>0$ then 
\eq
\Delta S^{(2)}_{\rm sing} = -\left[C\left({u\over {L^{x_{4s}}}}\right)^2+
D{{v}\over {L^{2+\ap}}}\right],
\ee 
where $C,D$ are non-universal functions of $TL^2$, whereas $v\equiv
g^2a^{2(2-\lambda)+2+\ap}$.
In this case a new scaling operator, fused from two interaction
operators, would emerge with a RG dimension $-(2+\ap)$. Moreover, 
since $2+\ap<2x_{4s}$ this operator would control the asymptotic scaling of 
$\Delta S_{\rm sing}^{(2)}$.
On the other hand, if $2x_{4s}-\ap-2<0$ the integral over the relative 
coordinates converges, and $\Delta S^{(2)}_{\rm sing} = 
-C({u\over {L^{x_{4s}}}})^2$, thus no new scaling variable needs to be 
introduced. Our results shown in Figure 2 indicate 
that $\ap=0$, thus  $2x_{4s}-2-\ap<0$ hence we do not need to introduce 
any new scaling operator at this order.

Now we summarize our results for short-range interaction. 
For interaction $V(r)=g/\vert r\vert^{\lambda}$,
we find that the non-interacting fixed point is {\it stable} 
(thus $z=2$ and $\nu\approx 2.3$) if $\lambda>2+x_{4s}$ 
(here $-x_{4s}$ is the RG dimension of short-range interactions). 
Our numerical results gives $x_{4s}\approx 0.65$.  
Although the above analyses do not form a ``proof'' that
strong short-ranged interactions are irrelevant,
we believe that the evidence is sufficiently strong.

Next, we consider the long-range Coulomb interaction, 
i.e., $\lambda=1$. In that case $\lambda < x_{4s}+2$, therefore
Eq.(8) is asymptotically controlled by 
$\Delta S^{(1)}_{\rm sing}=A{{g}/{L^{\lambda-2}}}$, 
which implies a relevant RG dimension for $g$ of $2-\lambda=1$. 
Thus the {\it non-interacting fixed point} is {\it unstable} 
upon turning on the Coulomb interaction. 
This result is not surprising given the fact that the measured value for $z$ 
is $1$ instead of the non-interacting value $2$. 
But if so, why should the static exponent $\nu$ remain unchanged? 

In two recent papers, MacDonald and coworkers studied the integer 
plateau transition under a Hartree-Fock treatment of the Coulomb 
interaction \cite{mac}. They found that a)
the DOS shows the Coulomb gap behavior
(i.e. $\rho(E_F)\sim L^{-1}$ in samples of linear dimension $L$
\cite{note10}) {\it  regardless of whether the system is at criticality
or not}; b) Despite the dramatic {\it non-critical} DOS suppression, 
the localization length exponent and the fractal dimension of the
critical eigen wavefunctions remain unchanged. 
In addition, the conductivities {\it did not} show any qualitative change. 
We take these results as indicating that 
the Hartree-Fock theory is in the same universality class 
as the non-interacting one. 
Thus the field theory should be the same nonlinear $\sigma$-model
\cite{llp} in which the {\it bare} parameters do not have non-trivial scale
dependence except that the DOS in the symmetry-breaking term should
be replaced by the appropriate Coulomb gap form.
%
Since the combination $\pi T\rho$ should have dimension $2$, 
and $\rho\sim 1/L$, it implies $z=1$. Thus, $z$ is modified while $\nu$
is not, and the change
in $z$ is caused by a {\it non-critical} modification of the DOS.

A direct consequence of the DOS suppression 
is that the dimension of $\psib_{\w_1}\psi_{\w_2}$ is changed from 0 to 1.
Indeed, it can be shown \cite{lw1} that
the two-particle spectral function
that is
consistent with the results in Ref.\cite{mac} and the scaling arguments
in Ref.\cite{chalker} is given by
\eq
S_2(E_1,E_2,{\vec q})={{\rho^2\s q^2}\over {\rho^2(E_1-E_2)^2+
(\s q^2)^2}}. 
\ee
In the above $\rho$ depends on $E\equiv(E_1+E_2)/2$ and $\s$, a quantity 
with the dimension of conductivity, depends on $\omega\equiv(E_1-E_2)/2$ 
and the wave vector ${\vec q}$. At the critical point,
$\rho(E)\sim 1/L$
and $\s(\omega,{\vec q})= {\rm  const.} $ for $\vert\rho\w\vert\gg q^2$; and 
${\rm const.}\times(q^2/\vert\rho\w\vert)^{x_2/2}$ for
$\vert\rho\w\vert\ll q^2$. Here $x_2\approx
-0.5$ is the exponent characterizing the anomalous diffusive behavior 
in the critical regime \cite{chalker,lw}. 
Note that the new exponents $x_{4s}$ is independent of $x_2$.
They are respectively the scaling dimensions of the operators associated
with the fusion products of four fermion operators, or two SU(2n) spin
operators that are symmetric and antisymmetric under permutations 
\cite{wegner}. If one uses Eq.~(13) to compute the two-particle Green's 
function, one can show that both $z$ and the scaling dimension 
of $\psib_{\w_1}\psi_{\w_2}$ are unity \cite{lw1}.

To support the predictions of the Hartree-Fock theory, one has to
analyze the stability of the Hartree-Fock fixed point when
the residual Coulomb interaction is taken into account.
Due to the normal ordering with respect to the Hartree-Fock ground
state, there is no contribution to $\Delta S_{\rm sing}^{(1)}$ 
due to the residual Coulomb interaction \cite{note}.
The lowest order effects now come in via $\Delta S_{\rm sing}^{(2)}$.
The new scaling form for $\Gamma^{(8)}$ is
$\Gamma^{(8)}(r_1,r_2,r_3,r_4,\{\w_i\},L)=L^{-4}{\cal F}_5
\left(r_{ij}L^{-1},\{\w_iL\}\right)$.
Inserting this result into
\eq
\Delta S^{(2)}_{\rm sing}=-{1\over {32}}({gT})^2{\sum_ {n_1...n_4}}'
\int\! d^2x d^2y d^2x' d^2y'{
\Gamma^{(8)}(x,y,x',y';\pi(2n_{1\to4}+1)T,L)
\over {\vert x-y\vert}{\vert x'-y'\vert}}
\ee
and {\it ignoring the possible short-distance divergence} we obtain 
$\Delta S_{\rm sing}^{(2)}\sim g^2(TL)^2$, thus $g$ is marginal. 
In order to go beyond this analysis 
(i.e. to determine the outcome of short-distance fusion) 
we need to know the behaviors of $\Gamma^{(8)}$ in a number of limits,  
information that we do not have at present.
Finally, we would like to emphasize that the Hartree-Fock 
theory presents a concrete example where, due to a {\it non-critical} 
suppression of the DOS, $z$ is modified while $\nu$ is not.

\noindent{\it Note added}. In an interesting recent work \cite{pb}, 
the effects of interactions are studied via the nonlinear $\sigma$-model
\cite{llp} where the topological term is handled by the dilute 
instanton gas approximation.
Since the latter {\it has not} produced the correct critical properties
even for the non-interacting transition, it is difficult for us
to judge the reliability of the results on the effects of interactions.

\noindent Acknowledgment: We thank J. Gan and S.~A. Kivelson for useful 
discussions.

\newpage
\bibliographystyle{unsrt}

\begin{thebibliography}{99}
\bibitem{note10}
We take this as indicating that in the thermodynamic limit 
$\rho(E)\propto\vert E-E_F\vert$.
\bibitem{note} We note that in Ref.\cite{mac} 
the Hartree-Fock is done for every disorder 
configuration. Hence here the normal ordering is performed 
before the impurity average.
\end{thebibliography}

\newpage
\centerline{\bf Figure Captions}

\ \

\noindent{\bf Fig.~1}. The scaling plot of $\Gamma^{(4)}$. Inset shows
the $L$-dependence of $\Gamma^{(4)}(x,x+1)$.

\ \

\noindent{\bf Fig.~2}. The scaling plot of
$\vert R-R^\prime\vert^{2x_{4s}}\Gamma^{(8)}$ obtained with
$x_{4s}\approx0.65$.
\vspace*{\fill}
\end{document}